\documentclass[multphys,vecphys]{promult}

\usepackage[]{amsmath}
\usepackage{makeidx}     
\usepackage{graphicx}    
\usepackage{multicol}    

\makeindex             

\begin{document}

\title*{Scheduling in targeted transient surveys and a new telescope
  for CHASE.}
\author{Francisco F\"orster\inst{1}, Nicol\'as L\'opez\inst{1}, Jos\'e
  Maza\inst{1}, Petr Kub\'anek\inst{2}, G. Pignata\inst{1}}
\institute{Departamento de Astronom\'ia, Universidad de Chile, Camino
  el Observatorio 1515, Las Condes, Chile \and Image Processing
  Laboratory, Universidad de Valencia, Pol\'igono La Coma s/n,
  Paterna, Valencia, E-46980, Spain.}
\maketitle

We present a method for scheduling observations in small
field--of--view transient targeted surveys. The method is based in
maximizing the probability of detection of transient events of a given
type and age since occurrence; it requires knowledge of the time since
the last observation for every observed field, the expected light curve
of the event and the expected rate of events in the fields where the
search is performed.

In order to test this scheduling strategy we use a modified version of
the genetic scheduler developed for the telescope control system
RTS2. In particular, we present example schedules designed for a
future 50 cm telescope that will expand the capabilities of the CHASE
survey, which aims to detect young supernova events in nearby
galaxies. We also include a brief description of the telescope and the
status of the project, which is expected to enter a commissioning
phase in 2010.

\section{Introduction}
\label{sec:intro}
With a new generation of observatories dedicated to studying the time
domain in astronomy \cite{PROMPT, Skymapper, CUMBRES, LSST}, our
understanding of astrophysical transient phenomena will be
significantly improved. The diversity of known families of transient
events will be better understood thanks to improved sample sizes and
better data, and new types of transient events will be likely
discovered.

These observatories will include large field--of--view, large aperture
telescopes, which will scan the sky in a relatively orderly fashion,
but also networks of small field--of--view, small aperture robotic
telescopes that will scan smaller areas of the sky in a less
predictable way.

The smaller robotic telescopes are ideal for studying very
short--lived transients, e.g. gamma ray bursts (GRBs), but also to do
detailed follow up studies of longer lived galactic (e.g. cataclysmic
variables, planetary systems) and extragalactic (e.g. supernovae)
transient events. Moreover, they constitute a relatively inexpensive
tool to obtain reduced cadences, of the order of days, in relatively
small areas of the sky which are of special interest, e.g. nearby
galaxies.

Here, we present a scheduling strategy that maximizes the probability
of finding specific types of transient phenomena, or the expected
number of events, at different times since occurrence. In
Section~\ref{sec:detecprob} we derive the probability of finding one
or more of these events, as well as the expected number of events. In
Section~\ref{sec:results} we show the results obtained with this
method and discuss its implications. Finally, in
Section~\ref{sec:50cm} we give an overview of the future 50 cm
telescope that will expand the capabilities of the CHASE survey and
which will use the scheduling method presented in this work.

\section{Detection probabilities of transient events}
\label{sec:detecprob}

This discussion will be limited to well known types of events in
targets with known distances. We assume that the light curves of every
transient event is composed of a monotonically increasing early
component, followed by a monotonically decreasing late component. We
will show how to compute detection probabilities for individual
difference observations, as well as for sequences of observations to
pre--defined targets. With this information, we will discuss how to
build observational plans that maximize the detection of events with
certain characteristics.

\paragraph{Expected numbers vs probabilities}

The probability of having exactly k occurrences of an event, in a time
interval where $\lambda$ occurrences are expected, is:
\begin{align}
  P(k, \lambda) = \frac{\lambda^k e^{-\lambda}}{k !},
\end{align}
which means that the probability of zero occurrences of the event is:
\begin{align} \label{eq:prob}
  P(0, \lambda) = e^{-\lambda},
\end{align}
which we would like to minimize.  Note that for small values of
$\lambda$, the probability of detecting at least one event should be a
better indicator of a good schedule than the total expected number, but
since $1 - e^{-\lambda} \approx 1 - (1 - \lambda) = \lambda$, in
practice this can be ignored. For big values of $\lambda$ the total
expected number of events should be most of the time a better
indicator of a good schedule than the probability of finding at least
one event.  For the purpose of this discussion we will use
probabilities, but it is easy to change the formulation to the
expected number of events, as we will show later.

\paragraph{Detection probabilities of individual events}

Let us assume that the events remain detectable for a time
$\tau$ and that their rate of occurrence is $R$. Consider also the case
when we look at a target twice to generate a difference image, with a
time interval or \emph{cadence}, $\Delta t$.

If each event remains visible for $\tau$ years, we would like to know
what is the interval where an event which was not seen in the first
observation could occur and be detected in the second
observation.

Let us also assume that the event was not seen in the first
observation, performed at time $t_1$, and that we make a second
observation with a cadence $\Delta t$, i.e. at time $t_2 = t_1 +
\Delta t$. Defining $\Delta t'$ as the minimum between $\Delta t$ and
$\tau$, then the time interval where new transients can occur and be
detectable in the second observation will span from $t_2 - \Delta t'$
and $t_2$. This is because short--lived transients only have a time
$\tau$ to remain visible, which could be smaller than $\Delta
t$. Hence, the expected number of new events that can be detected will
be the rate of occurrence times the former time interval. Using
equation~(\ref{eq:prob}), the probability of no events occurring in
this interval and no detections being made, $P_{\rm ND}$, will be:
\begin{align}
  P_{\rm ND} = \exp\lbrace -R~ \min(\Delta t,\  \tau) \rbrace.
\end{align}

\begin{figure}
\centering
\includegraphics[height=5cm]{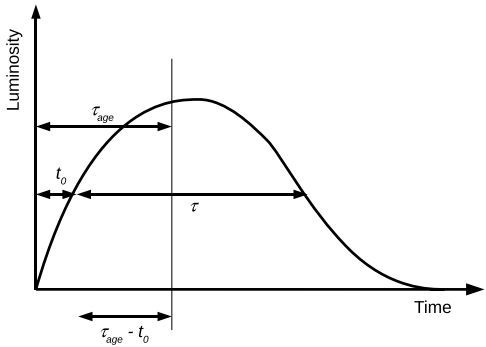}
\caption{The different variables used in this calculation. $t_0$ is
  defined as the time when the luminosity is above a certain threshold
  value which makes the object detectable, $\tau$ is defined as the
  time period where the transient can be detected and $\tau_{\rm age}$
  is an arbitrary age since explosion, with $\tau_{\rm age} - t_0$
  being the time period where events younger than $\tau_{\rm age}$ can
  be detected.}
\label{fig:def}
\end{figure}

Let us now assume that the event can only be detected a time $t_0$
after its occurrence, that it remains visible for a time $\tau$ and
that we are only interested in events younger than $\tau_{\rm age}$
(see Figure~\ref{fig:def}).

\begin{figure}
\centering
\includegraphics[height=5cm]{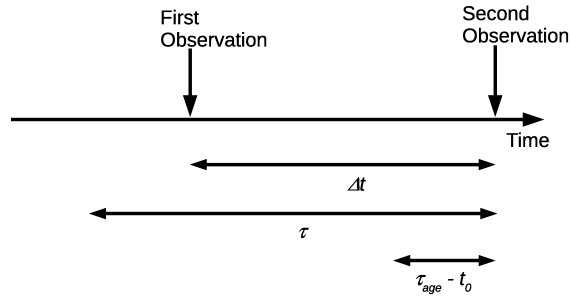}
\caption{Example of how the detection probabilities are
  calculated. $\Delta t$ is the cadence, $\tau$ is the time period
  when the transient remains detectable and $\tau_{\rm age} - t_0$ is
  the time period when the transient remains detectable while it is
  younger than $\tau_{\rm age}$. In this case, the cadence is shorter
  than $\tau$, which means that some transient events of the type of
  interest could have already been seen in the first observation, but
  $\tau_{\rm age} - t_0$ is shorter than the cadence, which means that
  all events younger than $\tau_{\rm age}$ could not have been seen in
  the first observation. The minimum between these three quantities
  should be multiplied by the rate of events in order to compute the
  expected number of events and compute the detection probabilities.}
\label{fig:example}
\end{figure}

The event will be detectable younger than $\tau_{\rm age}$ only if
$\tau_{\rm age} \ge t_0$. If this is the case, the time period where
events not seen in the first observation could occur and be detected
in the second observation will now be the minimum between $\Delta t$,
$\tau_{\rm age} - t_0$ and $\tau$ (see Figure~\ref{fig:example}).

Thus, the probability of no events occurring in this time interval and
no detections being made, $P_{\rm ND}^{\rm age}$, will be:
\begin{align}
P_{\rm ND}^{\rm age} = \exp \biggl[ -R ~\min \lbrace \Delta
  t,\ \max(\tau_{\rm age} - t_0,\ 0),\ \tau \rbrace \biggr].
\end{align}

With this information, the probability of detecting one or more events
in the second observation will be simply $1 - P_{\rm ND}^{\rm age}$.

\paragraph{Cadence choice}

Using the formula above, we could try maximizing the probability of
detection. For a fixed target, this can only be done decreasing the
cadence, $\Delta t$, as long as the number of targets that are
observed in the sample is not compromised significantly.

It is easy to see that, if $\tau_{\rm age} > t_0$, increasing $\Delta
t$ from zero to larger values will increase the probability of
detection only while $\Delta t < \min(\tau_{\rm age} - t_0,
\tau)$. For bigger cadences the probability will stay constant. Thus,
a natural choice for the cadence would be $\Delta t = \min(\tau_{\rm
  age} - t_0, \tau)$.

If larger cadences were chosen, the probability of detecting events
younger than $\tau_{\rm age}$ would not change, but not all detected
events would be guaranteed to be younger than $\tau_{\rm age}$, which
could be a problem if a fast age estimation is required. On the other
hand, larger cadences would increase the probability of detecting
older events, up to when $\Delta t > \tau$, when the probability of
detecting an event of any age would remain constant too.

Choosing $\Delta t = \min(\tau_{\rm age} - t_0,~ \tau)$ has the added
benefit that if $\tau_{\rm age}$ is smaller than the rise time and the
object type was known, the absolute magnitude of the event could be
used as an age estimator. This is because the event would be
guaranteed to be rising at detection time and the magnitude--age
relation would be single--valued. In reality, there could be events of
different types simultaneously occurring which would make the age
determination only useful in a statistical sense.

In general, $\Delta t$ will be a function of the distance and light
curve of the variable object to be detected, but also of the desired
age of detection, $\tau_{\rm age}$. Hence, for young and bright
objects with extended light curves, the cadence should be set to at
least $\tau_{\rm age} - t_0$ if we want to increase the cadence while
maximizing the detection of events with a given age.

However, it is not always easy to repeat the observations with a fixed
cadence. Bad weather, the change of position of the targets throughout
the year or the appearance of other objects of interest, among many
reasons, may cause the cadence between observations to vary.

An alternative strategy is to let the cadence adapt individually in a
sequence of observations in order to maximize the detection
probabilities.

\paragraph{Detection probabilities for a sequence of observations}

Now, we compute the probability of not detecting any new events in a
sequence of observations, $P_{\rm ND,~ Total}^{\rm age}$.

We note that for no events to be detected, each individual observation
must result in negative detections, i.e. we have:
\begin{align}
    P_{\rm ND,~ Total}^{\rm age} &= \Pi_i~ \exp \Bigl[ -R^i~ \min
    \lbrace \Delta t^i,\ \max(\tau_{\rm age} - t_0^i,\ 0),\ \tau^i
    \rbrace \Bigr] \notag \\ &= \exp \Bigl[ - \Sigma_i~ R^i~
    \min \lbrace \Delta t^i,\ \max(\tau_{\rm age} -
    t_0^i,\ 0),\ \tau^i \rbrace \Bigr],
\end{align}
where the indices indicate different targets. Thus, the probability of
detecting one or more new events will be:
\begin{align} \label{eq:probtotal}
    P_{\rm D,~ Total}^{\rm age} &= 1 - \exp \Bigl[ - \Sigma_i~
     R^i~ \min  \lbrace \Delta t^i,\ \max(\tau_{\rm age} -
      t_0^i,\ 0),\ \tau^i \rbrace \Bigr].
\end{align}
With this formula, we recover the expected number of events in the
entire observational sequence, which is the term inside the
exponential, i.e. in $P_{\rm D,~ Total}^{\rm age} = 1 - \exp(-
\lambda_{\rm D,~ Total}^{\rm age})$,
\begin{align} \label{eq:expected}
  \lambda_{\rm D,~ Total}^{\rm
  age} = \Sigma_i~ R^i~ \min \lbrace \Delta t^i,\ \max(\tau_{\rm age}
- t_0^i,\ 0),\ \tau^i \rbrace
\end{align} is the expected number of new detected
events with the required age. Thus, we can use either
equation~(\ref{eq:probtotal}) or (\ref{eq:expected}) to determine the
\emph{fitness} of individual schedules, but we recommend using
equation~(\ref{eq:expected}).

\paragraph{Limited number of targets}

It is possible that the number of targets available for detecting new
events with a given age is too small, i.e. assuming a fixed exposure
time and cadence for all observations, that the number of visible
targets where $\tau_{\rm age} > t_0$ is smaller than the length of the
night divided by the exposure time.

For very short--lived transient surveys this is not a problem, since
even with relatively small cadences $\Delta t > \tau$, and the probability of
detection in an individual observation would be $p \approx R~\tau$,
i.e. it would be almost independent of the cadence or how many times
we observe a target per night.

In relatively long--lived transient surveys, i.e. time--scales of days
or longer, we would not want to repeat targets in a given night. This
is because when $\Delta t < t_{\rm age} - t_0$ the probability of
detection in an individual observation is $P \approx R~\Delta
t$. Thus, many observations to a given target in a given night would
be almost equivalent to observing the target once per day or once
every few days in terms of probabilities, but with a significantly
higher cost on the resources and preventing the telescope from
observing other targets.

In general, the number of targets for a given cadence should be of the
order of the fraction of time that we want to spend in that sample per
night, $f$, times the total number of observations per night, $N_{\rm
  exp}$, times the cadence, $\Delta t$. The detection rate would be
approximately the multiplication of this number with the typical rate
of occurrence, $R$.

Hence, a possible strategy would be to order targets by the time that
it takes for the events of interest to be detectable, $t_0$, depending
on their distance and extinction, select a detection age according to
scientific criteria, and then group the targets according to the
resulting cadence and sample sizes. This is summarized in the
following table:

\begin{table}
\centering
\caption{For a desired detection age ($\tau_{\rm age}$) we show a
  possible choice of cadence in days ($\Delta t$), the maximum sample
  size consistent with this cadence and the approximate detection rate
  if the suggested cadence and maximum sample sizes were used. $t_0$
  corresponds to the time for the transient events to become visible
  since its occurrence, $N_{\rm exp}$ is the number of observations per
  night and $R$ is the rate of events expected in each field. We
  assume that in all cases the time that a transient remains visible,
  $\tau$, is bigger than the cadence chosen. For the opposite case,
  $\Delta t$ can be replaced by $\tau$ in the last two columns below:}
\label{tab:1}       
\begin{tabular}{c | c | c | c}
\ \ \ Age \ \ \  & \ \ Reference Cadence \ \ & \ \ \ Sample size \ \ \ & \ Approx. detection rate \ \\
\hline
$\tau_{\rm age}$ & $\Delta t = \langle \tau_{\rm age} - t_0 \rangle$ & $f~N_{\rm exp}~\Delta t$ 
& $f~N_{\rm exp}~ \Delta t ~\langle R \rangle$~\\
\end{tabular}
\end{table}

\vspace{.5cm}

\subsection{Genetic algorithm}

As discussed above, one can let the cadence vary from observation to
observation and from object to object. For an ideal schedule, we would
like to select the optimal combination of cadences that can adapt to
unexpected changes of the observational plan. For this, we use the
probability of detection, or the expected number of events, of a
sequence of observations as the \emph{fitness} indicator and we use a
genetic algorithm to find the best available observational plan for
the following night. This can reflect unexpected changes to the
observational plan in a daily basis, and can be extended to fractions
of a night optimizations if necessary.

We have used the genetic algorithm implemented in RTS2 \cite{rts2},
taking into account the cadence to each target ($\Delta t$) and the
distance, event rate, height above the horizon and sky brightness, all
of these reflected in the quantities $t_0$ and $\tau$, to build the
observational plan.

The distance between targets is also taken into account indirectly. If
it is too big, the number of visited targets per night, or the number
of terms in Equation~(\ref{eq:expected}), will be reduced and the
probability of detection will decrease accordingly. Similarly, when
the targets are too distant, or too close to the horizon, or the sky
too bright, $t_0$ will increase and $\tau$ will decrease, decreasing
the detection probabilities too. The bigger the event rate in every
target, the bigger the detection probability, which will favour those
targets with the biggest intrinsic rates. Finally, the time since last
observation will determine the cadence, changing the detection
probabilities as well.

In these calculations, the time between targets is computed using the
maximum between the slew time and the readout time, which effectively
defines a disk around each target where the time penalty is
constant. Reaching the outer circumference of this disk would take
exactly the readout time assuming that the CCD can read out electrons
while simultaneously slewing in the most efficient trajectory. This is
regularly accomplished by RTS2, since it optimizes observations by
reading out electrons and moving to a new position simultaneously.

For instance, a readout speed of about 2 sec and a slew speed of 5 deg
sec$^{-1}$ define a disk around the previous target of about 10 deg in
the sky where the time penalty for new targets is the same. In most
telescopes the slewing movement is accomplished with two independent
motors, which makes the size and shape of this disk really depend on
the initial configuration of the telescope before slewing, and whether
an equatorial or altazimuthal mount is used.

The details of the genetic optimizer, based on the NSGAII algorithm
\cite{nsgaII}, are described in detail in \cite{petrthesis}. It is
worth mentioning that the genetic algorithm can handle multiple
objectives, which can be used to find the Pareto front of optimal
values instead of a single solution, e.g. look for multiple detection
ages, which we have also implemented (see
Figure~\ref{fig:Paretofront}).

The Pareto front is the locus of solutions in a multi--objective
optimization problem where one objective cannot be improved without
compromising the other objective functions. For example, in an
optimization problem with two objective functions, for every value of
one of the two objective functions there is an optimal value for the
remaining objective function, i.e. the Pareto front can be composed by
infinite solutions.

\subsection{Calculation of $t_0$ and $\tau$}

In the previous sections we did not include the calculation of the
time for an event to become detectable, $t_0$, and the time that an
event remains detectable, $\tau$. These terms can be computed from
empirical light curves of the particular event to be detected, and can
be stored as functions of the critical luminosity above which the
object can be detected.

Thus, the problem is reduced to computing the flux above which the
object can be detected. To do this, we solve the signal to noise
equation for an arbitrary value above which we define an object to be
detected, e.g. $S/N = 5$. This equation is:
\begin{align} \label{eq:SN}
  S/N(t) = \frac{\gamma_{\rm TE}^{\rm CCD}(t)~ T}{\bigl[ \gamma_{\rm
        TE}^{\rm CCD}(t)~ T + \gamma_{\rm sky}^{\rm CCD}~ T~ n_{\rm pix} +
      \gamma_{\rm RN}^2~ n_{\rm pix} \bigr]^{1/2} },
\end{align}
where $S/N(t)$ is the signal to noise ratio as a function of time,
$\gamma_{\rm TE}^{\rm CCD}(t)$ are the photons per unit time detected
by the CCD from the transient event as a function of time, $T$ is the
exposure time, $\gamma_{\rm sky}^{\rm CCD}$ are the photons per unit
time coming from the sky and detected in one pixel of the CCD, $n_{\rm
  pix}$ is the number of pixels used to do photometry and $\gamma_{\rm
  RN}$ is the readout noise per pixel of the CCD. In general, $n_{\rm
  pix}$ is a function of the seeing at the zenith and the angle from
  the zenith.

Solving the previous quadratic equation for $\gamma_{\rm TE}^{\rm CCD}(t)$ with
a given value of $S/N$ and choosing the positive root gives the following
result:
\begin{align}
  \gamma_{\rm TE}^{\rm CCD}(t) = \frac{S/N^2}{2~T}~\Biggl[1 + \biggl( 1 +
    \frac{4~n_{\rm pix}}{S/N^2} \lbrace\gamma_{\rm RN}^2 + \gamma_{\rm
      sky}^{\rm CCD}~T \rbrace \biggr)^{1/2} \Biggr].
\end{align}
Now, we can include the effect of distance, collecting area, spectral
shape and CCD characteristics in the following equation:
\begin{align}
  \gamma_{\rm TE}^{\rm CCD}(t) = \frac{A}{4 \pi D^2}~\int \gamma_{\rm TE}^\nu(t)~\eta_\nu~d\nu
\end{align}
where $A$ is the collecting area of the telescope, $D$ is the distance
to the object, $\gamma_{\rm TE}^\nu(t)$ is the number of photons per
unit time per unit solid angle per unit frequency of the transient
event as a function of time since occurrence, $\eta_\nu$ is the
efficiency with which the photons are captured as a function of
frequency, which depends on the reflecting surfaces, intervening
lenses, CCD quantum efficiency and filters.

We can write a similar equation for the photons coming from the sky in
every pixel of the CCD:
\begin{align}
  \gamma_{\rm SKY}^{\rm CCD}(t) = \frac{\Delta\Omega~A}{\cos z}~\int \gamma_{\rm
    SKY}^\nu(t)~\eta_\nu~d\nu,
\end{align}
where $\Delta \Omega$ is the solid angle of one pixel of the CCD, $z$
is the angle from the zenith and $\gamma_{\rm SKY}^\nu$ is now the
number of photons coming from the sky per unit time per unit area per
unit solid angle per unit frequency.

Thus, if we compute $\gamma_{{\rm TE}}^{\rm CCD}(t)$, assuming the
object is at a distance $D$ from the observer, $\gamma_{{\rm
    TE},D}^{\rm CCD}(t)$, and $\gamma_{\rm SKY}^{\rm CCD}$ assuming
the object is at a given angle $z$ from the zenith, $\gamma_{\rm
  SKY}^{\rm CCD}(z)$, for a given sky brightness and for a particular
telescope configuration, we can simply scale the results as follows:
\begin{align}
  \gamma_{{\rm TE},D}^{\rm CCD}(t) = \gamma_{{\rm TE},D_0}^{\rm
    CCD}(t)~\biggl(\frac{D_0}{D}\biggr)^2 \label{eq:TED}\\
  \gamma_{\rm SKY}^{\rm
    CCD}(z) = \frac{\gamma_{\rm SKY}^{\rm CCD}(0)}{\cos z},
\end{align}
Thus, we can now compute the times when the object becomes detectable
and when it is no longer detectable, $t_0$ and $t_0+\tau$:
\begin{align}
  t_0,&\ t_0+\tau = \notag \\
  &\biggl(\gamma_{{\rm TE}, D_0}^{{\rm CCD}}\biggr)^{-1} \Biggl\lbrace
  \biggl(\frac{D}{D_0}\biggr)^2 \frac{S/N^2}{2~T}~\Biggl[1 + \biggl( 1
    + \frac{4~n_{\rm pix}}{S/N^2} \lbrace\gamma_{\rm RN}^2 +
    \frac{\gamma_{\rm sky}^{\rm CCD}(0)}{\cos z}~T \rbrace
    \biggr)^{1/2} \Biggr] \Biggr\rbrace,
\end{align}
where $\biggl(\gamma_{{\rm TE}, D_0}^{{\rm CCD}}\biggr)^{-1}$ is the
inverse of the function computed in equation~(\ref{eq:TED}), which
should have two solutions for a transient which is composed by an
early monotonically increasing component followed by a monotonically
decreasing late component. Importantly, the inversion of $\gamma_{{\rm
    TE}, D_0}^{{\rm CCD}}$ must be performed only once, and can be
stored numerically in a table, e.g. in logarithmic intervals of
photons per unit time.

Thus, for a given signal to noise ratio ($S/N$), which we arbitrarily
define as the value that gives a detection, a given distance from the
source ($D$), exposure time ($T$), sky brightness (c.f. $\gamma_{\rm
  SKY}^{\rm CCD}$), seeing (c.f. $n_{\rm npix}$), readout noise per
pixel ($\gamma_{\rm RN}$) and angle from the zenith ($z$), we can
compute $t_0$ and $\tau$, which are necessary for the calculation of
the expected number of events and the probabilities of detection in an
individual target and a sequence of observations.

It is important to note that the detection of objects is sometimes
performed using individual pixels, in which case we can set $n_{\rm
  pix}$ to one, and multiply the term $\gamma_{\rm TE}^{\rm CCD}(t)$
in equation~(\ref{eq:SN}) by the fraction of photons that fall in the
central pixel in the position of the object, depending on the seeing
conditions, which would result in the following modified equation:
\begin{align}
  t_0,&\ t_0+\tau = \notag \\
  &\biggl(\gamma_{{\rm TE}, D_0}^{{\rm CCD}}\biggr)^{-1} \Biggl\lbrace
  \biggl(\frac{D}{D_0}\biggr)^2 \frac{S/N^2}{2~f~T}~\Biggl[1 + \biggl( 1
    + \frac{4}{S/N^2} \lbrace\gamma_{\rm RN}^2 +
    \frac{\gamma_{\rm sky}^{\rm CCD}(0)}{\cos z}~T \rbrace
    \biggr)^{1/2} \Biggr] \Biggr\rbrace,
\end{align}
where $f$ is the fraction of the light from a point source that would
fall in one pixel in the position of the object, generally a function
of the seeing at the vertical, the angle from the zenith and the
frequency of the photons to be detected.

\section{Results and discussion} \label{sec:results}

\begin{figure}
\centering
\includegraphics[height=9cm]{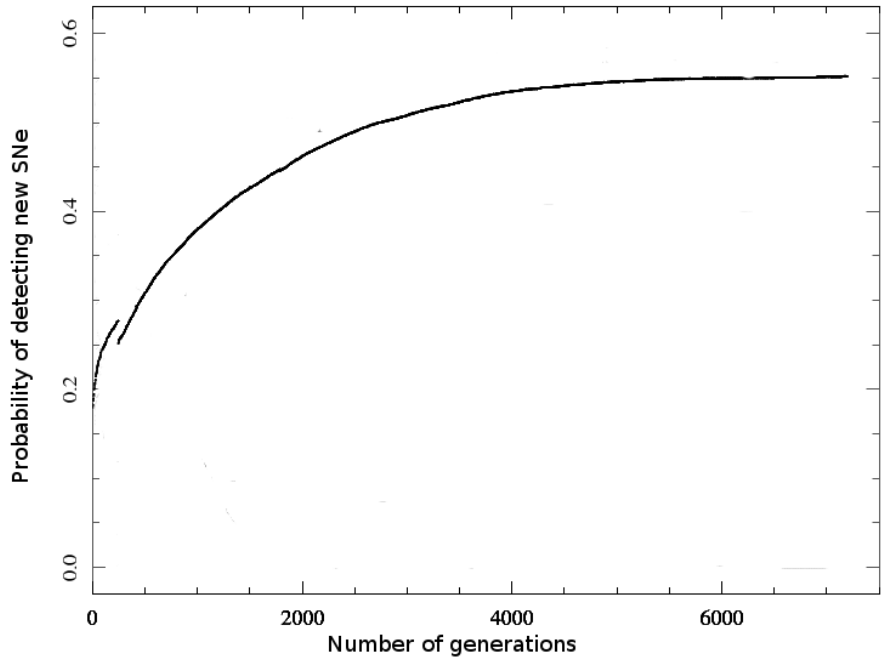}
\caption{Evolution of the probability of detecting new supernova with
  the number of generations. In this example, every generation
  consists of a population of 1,000 observational plans, where each
  observational plan contains hundreds of 60 sec exposures to
  different targets. For the simulation, we have used the gold sample
  of galaxies of the CHASE survey \cite{CHASE}, where the distances
  were computed from recession velocities. Event rates were created
  randomly at the beginning of the simulation to reproduce the typical
  rates expected for supernova explosions. The cadences were randomly
  generated to reproduce the typical values expected in the CHASE
  survey.}
\label{fig:generations}
\end{figure}

\begin{figure}
\centering
\includegraphics[height=9cm]{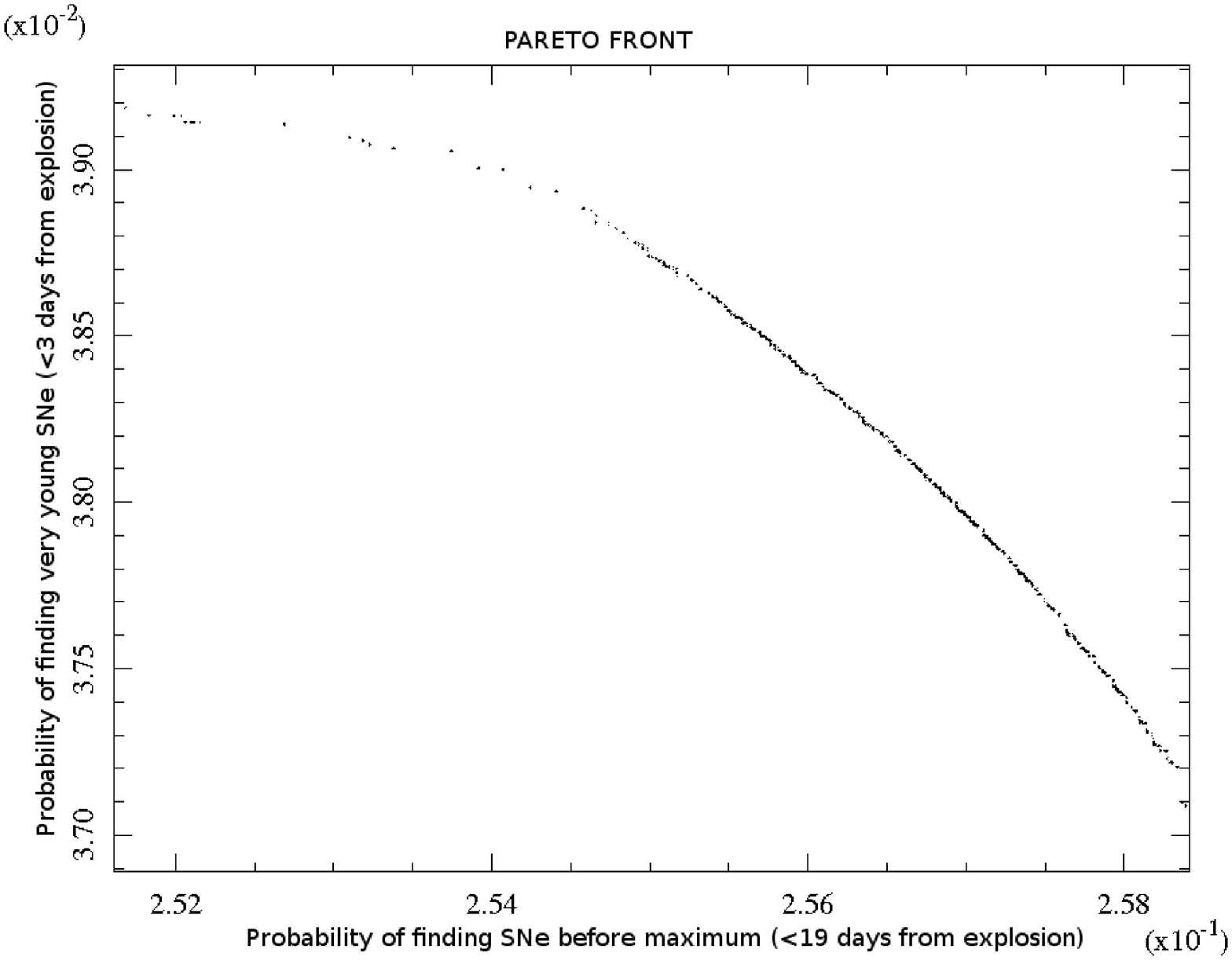}
\caption{Locus of Pareto--optimal solutions, or Pareto front, using
  two objective functions: the probability of finding SNe before
  maximum (abscissa) and the probability of finding SNe no later than
  three days after the explosion (ordinate). In this example we have
  evolved 10,000 generations of 1,000 observational plans each,
  similar to the simulation shown in Figure~\ref{fig:generations}, but
  with two objectives instead of only one. Absolute numbers should not
  be trusted, since the simulated cadences are too big for the values
  expected in a survey looking for SNe as young as three days after
  explosion. Individual points represent individual observational
  plans, and the set of points are only an approximation to the Pareto
  front. Once the Pareto front is computed, one observational plan
  from the set of solutions can be chosen according to arbitrarily
  defined criteria.}
\label{fig:Paretofront}
\end{figure}

In Figures~\ref{fig:generations} and \ref{fig:Paretofront} we present
example implementations of the scheduling strategy presented in this
work with the genetic algorithm used in RTS2.

Figure~\ref{fig:generations} shows the probability of finding
supernova with a reference 50 cm telescope in an observational plan
composed of 60 sec individual exposures, with simulated cadences and
supernova rates in each field. We can see the probability increasing
with each generation of observational plans and then staying
constant. Each generation is formed by a population of 1,000 different
observational plans and the initial iteration consisted of a series of
randomly generated targets for each observational plan of the
population, which were crossed and mutated to obtain the best
observational plans.

Figure~\ref{fig:Paretofront} shows the space of optimal solutions when
two objective functions are used. This is, the Pareto front of
non--dominated solutions, or the space of solutions where one variable
is at its optimal value without compromising the other variables. In
this simulation we use the objective functions: (1) probability of
finding supernovae before maximum and (2) the probability of finding
supernova younger than three days from explosion, using similar
parameters to those used in the simulation shown in
Figure~\ref{fig:generations}.
 
Interestingly, we have used the already implemented genetic scheduler
from RTS2 to find the schedule that maximizes the average height above
the horizon for our list of targets, or that minimizes the typical
distance between targets. For both cases, we have found that the
probability of detection of the resulting schedule is smaller by more than a
factor of two with respect to our method, which suggests that our
strategy is significantly better for finding transient objects.

Thus, the implemented scheduling strategy based on maximizing the
probability of finding new transient events is able to obtain
significantly higher detection probabilities than alternative
methods. We were able to build observational plans for every night to
maximize the probability of detecting particular events, or similarly,
the expected number of detections. These plans were based on
pre--defined samples of targets that have characteristic cadence and
exposure times, and that can easily adapt to unforeseen changes in the
scheduled observations.

In order to compute the observational plans with the highest detection
probabilities, we used the genetic algorithm implemented in the
telescope control system RTS2, where a
multi--objective algorithm selects the optimal sequence of
observations for our purposes.

We expect to be able to extend this work to scheduling of coordinated
networks of robotic telescopes looking for specific types of transient
events, or looking for many different phenomena if multi--objective
optimization is used. We also expect to release the implementation in
a future version of RTS2  (\verb+http://rts2.org+).

An important question is whether this method is able to recompute the
optimal observational plan when unexpected changes in the sequence of
observations occur. In a single computer, with the current
implementation of the code we cannot think of simple ways of achieving
this, since it normally takes many hours to find the optimal
observational plan or set of Pareto--optimal plans in a single
PC. However, with faster computers, pre--calculating detection
probabilities for every target at every time in the night, and given
that genetic algorithms can be relatively easily parallelised, we
expect this to be feasible in the near--future.

Alternatively, one could switch from using optimized observational
plans to computing the detection probabilities for every available
target and choose the one with the highest detection probability every
time the telescope has finished integrating, taking into account the
slew and readout time by subtracting the expected cost of slewing in
term of detection probabilities per unit time for the corresponding
slewing times.

Finally, it should be noted that this method is not exclusive for
supernova transients, but to any transient with well characterized
light--curves and with well understood target fields.

\section{Application to the new 50 cm robotic telescope for CHASE} \label{sec:50cm}

\begin{figure}
\centering
\vbox{
\includegraphics[width=0.5\hsize]{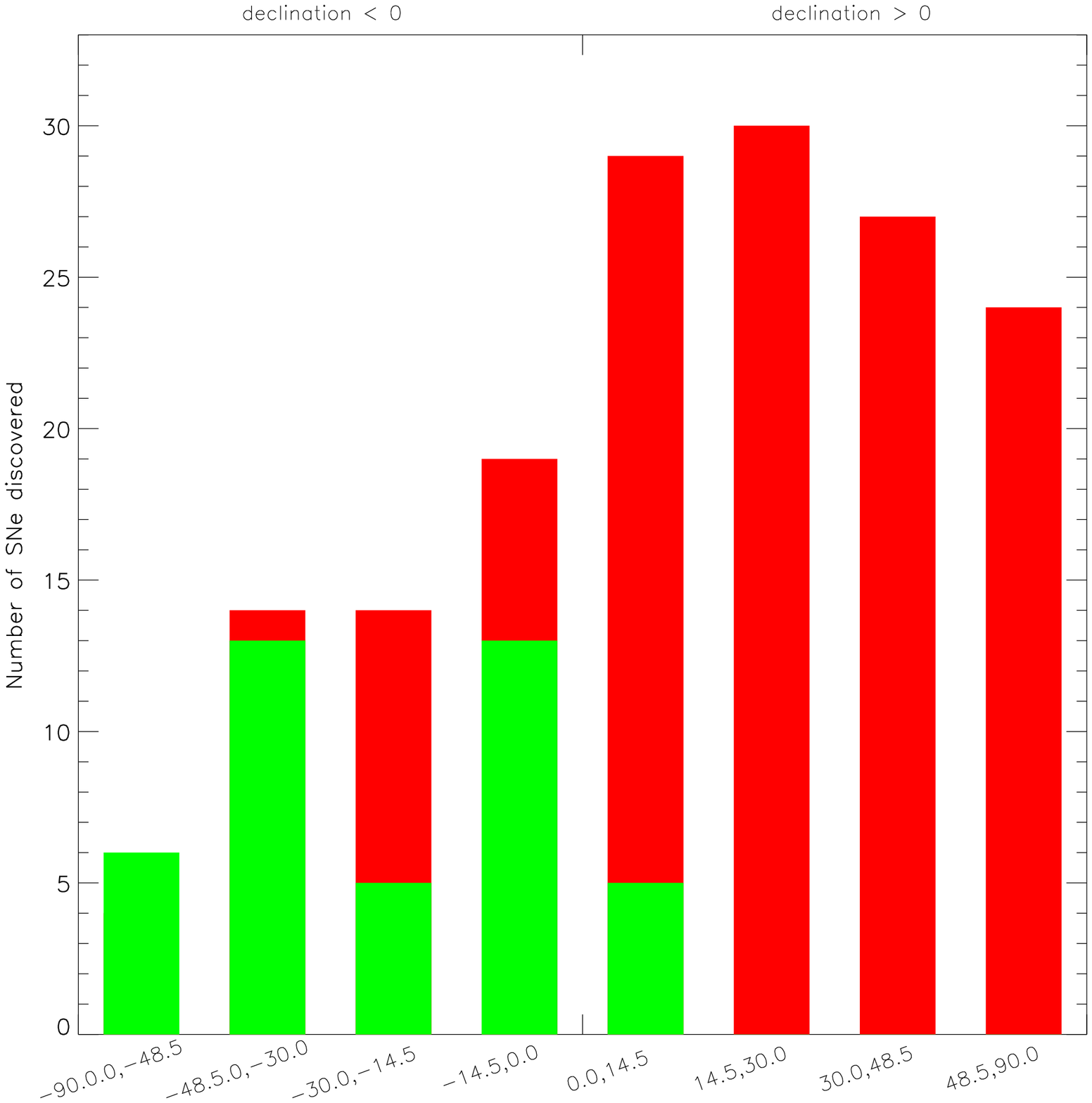}
}
\vbox{
\includegraphics[width=0.49\hsize]{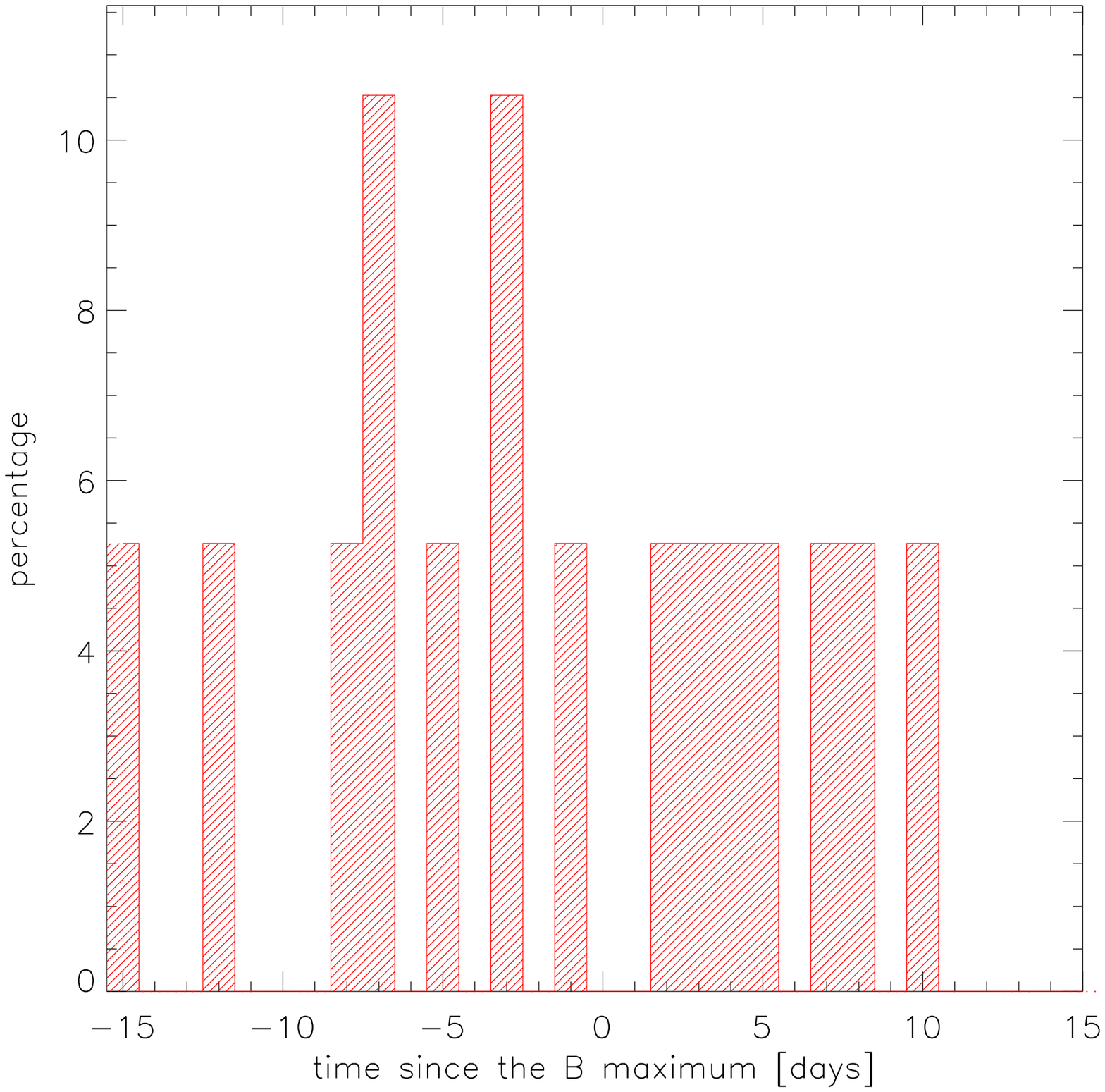}
\includegraphics[width=0.49\hsize]{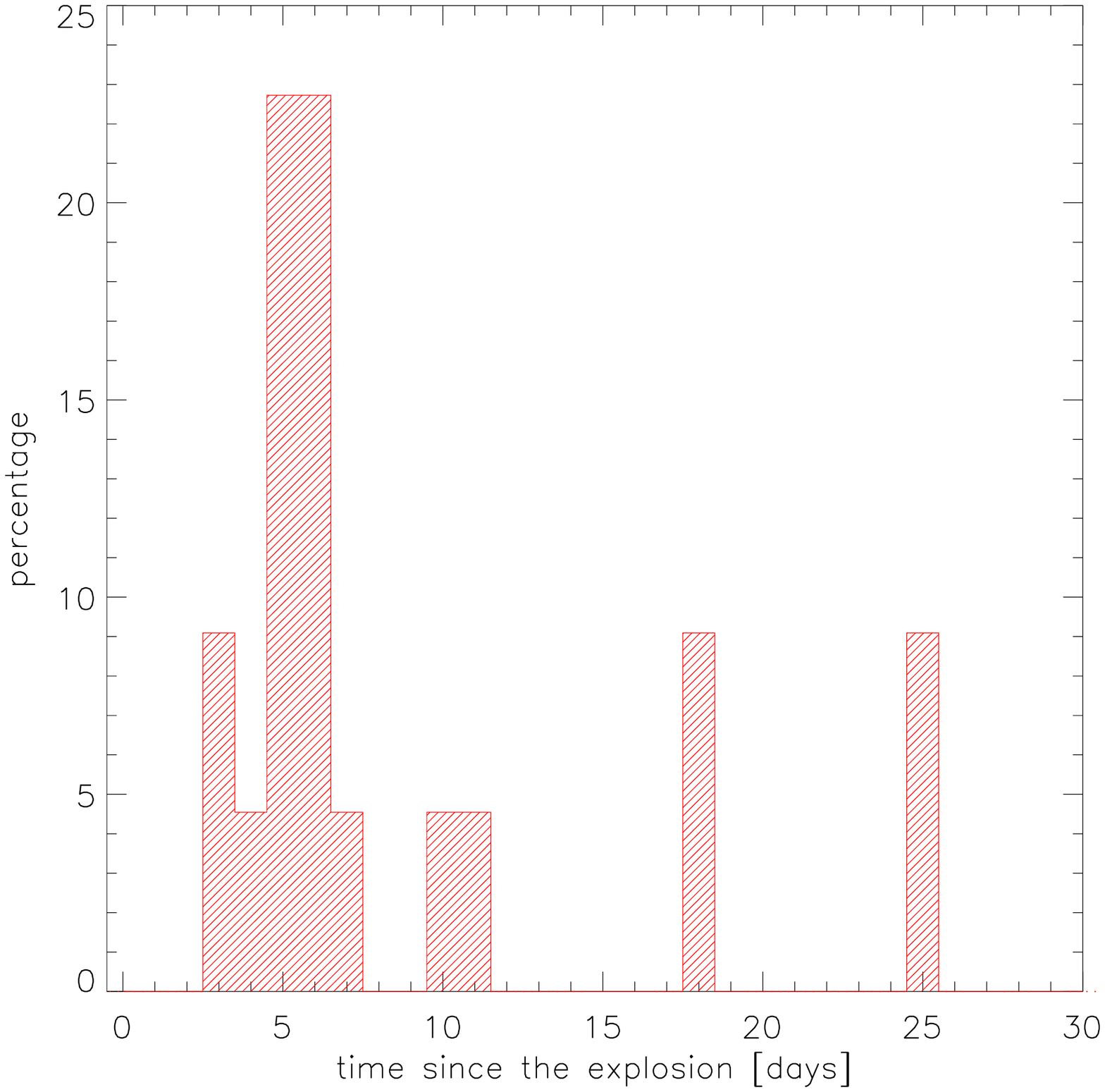}
}
\caption{CHASE survey results. \emph{Top:} the distribution of
  supernova with declination in bins of equal solid angle. Red bins
  correspond to all nearby SNe found in 2008, and green bins
  correspond to SNe discovered by CHASE in 2008. \emph{Bottom--left:}
  the distribution of discovery ages of Type Ia SNe in CHASE. About
  half of the SNe were discovered before
  maximum. \emph{Bottom--right:} the distribution of discovery ages of
  Type II SNe in CHASE. The median discovery age was about 5 days
  after explosion. (G. Pignata, private communication).}
\label{fig:chase}
\end{figure}

The CHASE survey \cite{CHASE} is the most prolific nearby supernova
search in the southern hemisphere. It finds more than 70\% of the
nearby ($z < 0.3$) supernova in the southern hemisphere, with
discovery ages much younger than competing surveys (see
Figure~\ref{fig:chase}). CHASE uses a fraction of the time available
in four of the six PROMPT telescopes \cite{PROMPT} located in CTIO.

In order to expand the capabilities of CHASE and to have a better
control over the scheduling of the observations, we are in the process
of purchasing and installing a 50 cm robotic telescope that will join
the other PROMPT telescopes for the SN survey and follow up.

The telescope will be a 50 cm automated telescope: composed of an
optical tube, a CCD camera with a set of filters, a mount, a
meteorological station, a dome and computers for controlling and
analyzing the data. It will be located in CTIO and remotely controlled
from Cerro Cal\'an (Santiago, Chile). It will observe hundreds of
targets every night with the aim of doubling the observing
capabilities of the CHASE survey and to try new observing strategies
with new associated scientific goals.

\begin{figure}
\centering
\includegraphics[width=\hsize]{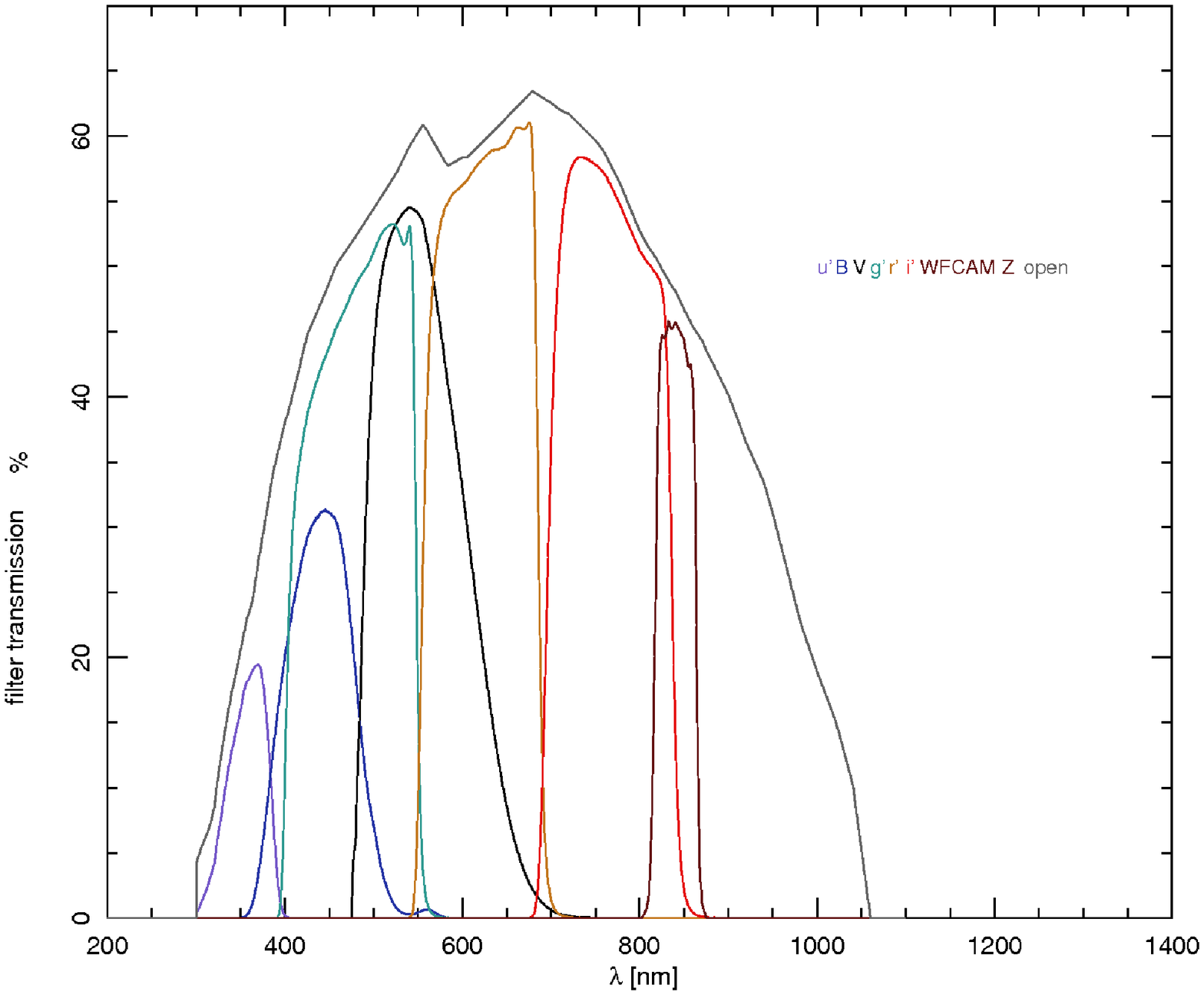}
\caption{Approximate transmission curves for the filters implemented
  in our future 50 cm telescope. ``Open'' corresponds to no filters,
  and includes the effects of atmospheric extinction, mirror
  reflectivity and CCD quantum efficiency. The supernova search will
  be likely performed in ``Open'' mode or with a clear filter, but for
  the follow--up program we will use the filters shown here. It is
  possible that in the future we will include additional filters.}
\label{fig:filters}
\end{figure}

The optical tube of the telescope will be a 50 cm aperture
Ritchey--Chretien design, with a focal ratio of 12, in an open--truss
carbon fiber tube purchased from the Italian company Astrotech. The
camera will be a 2kx2k pixels Finger Lakes Proline camera, with a back
illuminated, UV enhanced, 95\% peak quantum efficiency Fairchild 3041
CCD.  The pixel size will be 0.52$''$ and the field--of--view will be
17.6$'$ in side. The camera will be equipped with a 12--slot filter
wheel with the filters u'g'r'i', Johnson B and V and WFCAM Z,
purchased from Asahi--Spectra (see transmission curves in
Figure~\ref{fig:filters}).

The camera was chosen to avoid the potential presence of residual
images in the imaging of targets, which currently dominate our SN
candidate lists with the PROMPT telescopes, to obtain a relatively big
field--of--view, which would allow us to image enough reference stars
to do an accurate image alignment and subtraction, but also to obtain
the best available quantum efficiency, which is a cost--effective way
of collecting more photons per target.

The mount will be the Astro--Physics 3600GTO ``El Capit\'an'' model,
which is a German equatorial mount with sub-arcmin pointing errors,
and a slew speed of about 5 $\deg\ \sec^{-1}$. The dome of the
telescope will be built in Chile and is currently in the design phase.

The scheduling of the observations will be done with the strategy
presented in this work, and we expect to start collaborations with
other groups using this scheduler in an integrated fashion. For more
information please contact the authors.

\paragraph
\newline
{\it Acknowledgment.}  We acknowledge an anonymous referee whose help
and guidance lead to significant improvements to the
manuscript. F.F. acknowledges partial support from GEMINI-CONICYT
FUND. G.P.  acknowledges partial support from the Millennium Center
for Supernova Science through grant P06-045-F funded by ``Programa
Bicentenario de Ciencia y Tecnolog\'ia de CONICYT'' and ``Programa
Iniciativa Cient\'ifica Milenio de MIDEPLAN''.
\end{document}